%% file: main_technical_report.tex
\documentclass[onecolumn,letterpaper, 10 pt, conference]{ieeeconf}  

\IEEEoverridecommandlockouts                              

\usepackage{graphicx} 
\usepackage{amstext}
\usepackage{cite}

\title{Technical report \\ {\LARGE \bf
Hierarchical Robust Analysis for Identified Systems in Network}}

\author{Anton Korniienko$^{1}$ and Xavier Bombois$^{1}$ and H\aa kan Hjalmarsson $^{2}$ and G\'erard Scorletti$^{1}$
\thanks{$^{1}$Anton Korniienko, Xavier Bombois and G\'erard Scorletti are with Laboratoire Amp\`{e}re, Universit\'{e} de Lyon, \'{E}cole Centrale de Lyon, 69134 Ecully Cedex, France, {\tt\small first\_name.last\_name@ec-lyon.fr}.}
\thanks{$^{2}$ H\aa kan Hjalmarsson Automatic Control, School of Electrical Engineering, KTH, 100 44 Stockholm, Sweden,
        {\tt\small hakan.hjalmarsson@ee.kth.se}}%
}

\input{setup.tex}

\begin{document}

\maketitle
\thispagestyle{empty}
\pagestyle{empty}

\begin{abstract}

This technical report considers worst-case robustness analysis of a network of locally controlled uncertain systems with uncertain parameter vectors belonging to the ellipsoid sets found by identification procedures. In order to deal with computational complexity of large-scale systems, an hierarchical robustness analysis approach is adapted to these uncertain parameter vectors thus addressing the trade-off between the computation time and the conservatism of the obtained result.

\end{abstract}


\section{Introduction}

In this technical report, the problem of worst-case robustness analysis of a network of locally controlled uncertain Linear Time Invariant (LTI) subsystems is under consideration. The uncertainty of each subsystem is an uncertain real vector that belongs to an ellipsoid: an uncertainty set in the model parameter space typically obtained after identification.  

This work is motivated by recent technological advances in Microelectronics, Computer Sciences, Robotics, and related topics in the field of the Multi-Agent systems~\cite{CYRC:13}. The control of these network systems is usually decentralized and in order to compute controllers achieving high performance level, the model of the subsystems needs to be known. An efficient method to build the appropriate models is system identification~\cite{ljung1999system}. However, due to the presence of the noise and since the identification experiment is limited in time, the model parameters can only be identified within some prescribed uncertainty region which is typically an ellipsoid. For these reasons, in order to ensure that the computed controllers achieve the performance not only for the nominal identified model but for the true network system, it is important to take into account these uncertainties. The evaluation of the uncertainty effects on the system stability and performance is called robustness analysis.

The large scale of today's systems raises additional challenges on identification, controller design as well as on the robustness analysis. In this technical report we focus on the robustness analysis in the context of large-scale network systems.

In the 80's-90's, $\mu$-analysis \cite{Doy:82,Saf:82} was developed to investigate the performance of LTI systems in the presence of structured uncertainties. The performance is evaluated in the frequency domain~\cite{SkP:05}. This approach is based on the computation of the structured singular value $\mu$ of the frequency dependent matrices, which was proved to be NP-hard \cite{BYDM:94}. Fortunately, lower and upper bounds on $\mu$ can be efficiently computed; the $\mu$ upper bounds in \cite{FTD:91} guarantee a certain level of performance with some conservatism. By efficient, it is understood that the computation time is bounded by a polynomial function of the problem size \cite{GaJ:79}. An adaptation of these results to classes of the uncertainties obtained by identification can be found in~\cite{BGSA:01,SBBF:07,BBHS:08}.

Nevertheless, even if the computation of the $\mu$ upper bound is efficient, its computation time can be important in the case of uncertain large-scale systems. The purpose of this technical report is to extend the results~\cite{BGSA:01,SBBF:07,BBHS:08} to the context of large-scale interconnected systems, addressing the trade-off between computation time and conservatism. To do so, we adapt the hierarchical robustness analysis approach of~\cite{DKS:13,DKS:14,LKSM:15}, initially proposed in~\cite{Saf:83b}, to the class of uncertainties obtained from system identification. A similar subject is presented in our current work~\cite{BKHS:17}. The main contribution of this technical report is, however, a deeper investigation of the robustness analysis aspects, allowing, in contrast to~\cite{BKHS:17}, for several types of embedding and their combinations.

The next section of the technical report formulates the problem under consideration, while the third section presents the main result of the technical report, the hierarchical analysis approach.  The fifth section is dedicated to the numerical illustration example and the last section concludes the technical report. Below we give notation used in the sequel.

\paragraph*{Notations}

We denote by $H \star M$ the transfer function
$
 M_{22}+M_{21}H \left(I-M_{11}H \right)^{-1}M_{12}
$
with $M_{ij}$ being appropriate partitions of $M$ and~$\star$~standing for the Redheffer star product: it will be referred to as the  Linear Fractional Transformation (LFT) interconnection of $M$~and~$H$.
The matrix \[ \BA{ccc} X_1 & 0 & 0\\ 0 & \ddots & 0 \\ 0 & 0 & X_N \EA \] is denoted as $ \dia_{i} (X_i)$ with (block-)diagonal elements $X_i$ ($i=1,...,N$). 
For a complex number $y$, we denote $yy^*$ by $y^2$ while $\bar \sigma(A)$ denotes the maximal singular value of a complex matrix $A$.

\section{Problem Statement}

Let us consider a network of $\Nm$ single-input single-output (SISO) subsystems $\cS_i$ ($i=1...\Nm$) operated in closed loop with a SISO decentralized controller $K_i$ ($i=1...\Nm$):
\vspace*{-0.0cm}
\be \label{module} \cS_i(\theta_i): \ y_i(t)=G_i(\sz ,\theta_i) u_i(t) + v_i(t)  \ee 
\vspace{-0.5cm} \be \label{excit1} u_i(t) = K_i(\sz)(r_{i}(t)-y_i(t))  \ee 
\vspace*{-0.5cm}\be   \label{interco}  \bar r(t) = \cA \ \bar y(t) \ + \ \cB \ ref(t) 
\ee where $\sz$, in order to keep the discussion as general as possible and to consider both cases, defines the Laplace variable $s$ in the continuous time domain or the shift variable $z$ in the discrete time domain. The vector $\theta_i \in\reals^{n_{\theta_i}}$ represents the parameter vector of the $i$th system. We will distinguish hereafter between a variable $\theta_i \in\reals^{n_{\theta_i}}$, its unknown true value, $\tio \in\reals^{n_{\theta_i}}$, and its estimated value, $\thi \in\reals^{n_{\theta_i}}$. Let us also define $\theta = \left[\theta_{1},\dots,	\theta_{N}\right]^T \in\reals^{n_{\theta}}$, $\to \in\reals^{n_{\theta}}$ and $\th \in\reals^{n_{\theta}}$: the stacked version of the previous parameter vectors, with $n_{\theta} = \sum_i n_{\theta_i}$. The signal $u_i$ is the input applied to the system $\cS_i$ and $y_i$ is the measured output. This output is made up of a contribution of the input $u_i$ and of a disturbance term $v_i$ that represents both process and measurement noises and is modeled as a stochastic random process~\cite{ljung1999system}. The different true systems are thus described by transfer functions $G_i(\sz,\tio)$. Moreover, the vector $\bar v \eqbydef (v_1 , v_2 ,...,v_{\Nm})^T$ is assumed to have mutually independent components $v_i$.

The subsystems $\cS_i (\tio)$ in~(\ref{module}) may all represent the same type of subsystems combined into the network in order to achieve some global goals. Due to industrial dispersions, the unknown parameter vectors $\tio$ may, of course, be different for each $i$, the same applies to the order of the transfer functions $G_i$. 

In this technical report, the interconnection form used in formation control or multi-agent systems (see e.g. \cite{FaM:04,KSCB:16}) is under consideration. Each subsystem $\cS_i (\theta_i)$ is operated with a decentralized controller $K_i(\sz)$, see~(\ref{excit1}), and the signal $r_{i}$ is a locally available reference signal that will be computed via~(\ref{interco}). The matrices $\cA$ and $\cB$ in~(\ref{interco}) represent the interconnection (flow of information) physically present in the network. Furthermore, $\bar r$, $\bar y$ are defined in the same way as $\bar v$ above. 
A possible main global objective of the network could be the tracking performance: each output $y_i(t)$ has to approach in a specified time the reference signal: $ref_i(t) = ref(t)$. However, the external reference signal $ref(t)$ is generally only available (throughout $r_i$) at one or a few nodes of the network, which is defined by the matrix $\mathcal{B}$. 

\begin{figure}
 	\centering
 	\includegraphics[width=180pt]{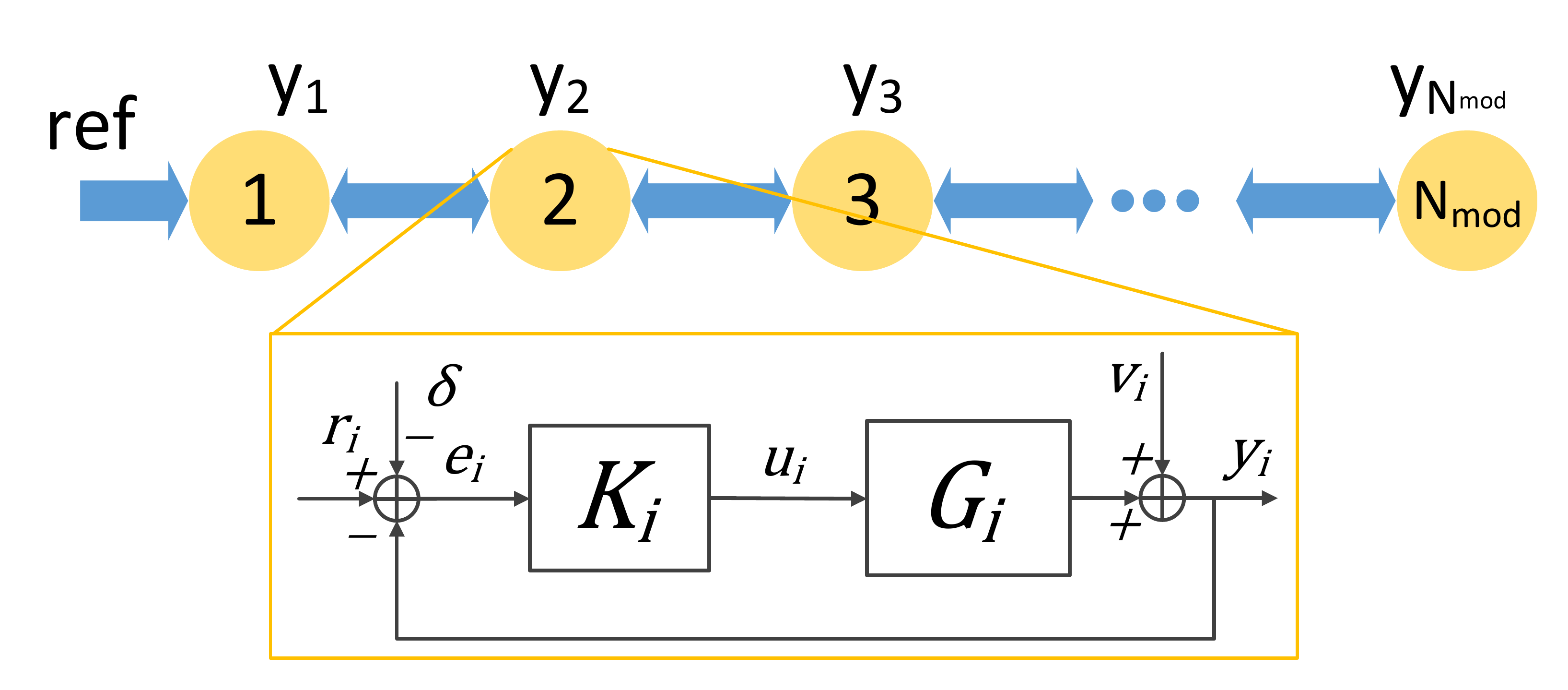}
 	\caption{Example of a network}
 	\label{figexnet} \vspace*{0cm}
 \end{figure}

As an example, let us consider the network in Fig.~\ref{figexnet} (consider $\delta=0$ for this part) with $\Nm$ systems connected in a chain, all of the form~(\ref{module}) and all with a decentralized controller $K_i$, see~(\ref{excit1}). These local closed loops are represented by a circle and are detailed in Fig.~\ref{figexnet}. In order to be able to track the external reference $ref$ even though this reference is only available at Node $1$, a number of nodes are allowed to exchange information (i.e. their measured outputs) with some other neighboring nodes. The arrows between the nodes in Figure~\ref{figexnet} indicate the flow of information. For example, Node $2$ sends its output to/receives the outputs from Nodes $1$ and $3$ while Node $1$ receives the output of Node $2$ and from the external reference signal and sends its output only to Node 2. The local reference signal $r_{i}$ of Node $i$ will be computed as a linear combination of the received information at Node $i$. 
More precisely, to define all outputs $y_i$, $\cA$ and $\cB$ in~(\ref{interco}) are chosen as \cite{FaM:04,KSCB:16}:  

{\small \label{eq:AB}\be\cA=\BA{ccccc} 0 & 1/2 & 0 & \cdots & 0 \\  1/2 & 0 & 1/2 & \cdots & 0\\ \vdots & \ddots & \ddots & \ddots & 0 \\ 0 & 0 & 1/2 & 0 & 1/2\\  0 & 0 & 0 & 1 & 0  \EA \ \ \ \cB = \BA{c}1/2\\0\\\vdots\\0\EA. \ee}
 
\ni The matrix $\cA$ is called the normalized adjacency matrix in the literature \cite{FaM:04} and it can be easily obtained for any interconnection topologies. Using~(\ref{interco}), it is possible to define the local tracking error signals $e_i = r_{i}-y_i$ and it can be proven that such an interconnection allows good tracking if different loops $[K_i \ G_i]$ are designed to make the tracking error $e_i$ as small as possible. Our objective is thus to design (or redesign) local controllers $K_i$ ensuring this global objective for a given interconnection topology $\mathcal{A}$, $\mathcal{B}$ and given subsystem dynamics $G_i(\sz ,\tio)$, see~(\ref{module})-(\ref{interco}).

Let us first define general performance specifications that cover the expressed tracking performance objective but also other additional specifications. To do so, let us introduce performance input $\bar w$ and output $\bar z$ and a (possible dynamic) interconnection matrix $\cM$ such that
\vspace*{-0.0cm}\be   \label{intercoM} 
\begin{array}{ccc}
	\left[\begin{array}{c}
		\bar r\\
		\bar z
	\end{array}\right] & = \ \cM & \left[\begin{array}{c}
		\bar y\\
		\bar w
	\end{array}\right]\end{array} 
\ee 
Different components of the matrix $\cM$ depend on the information flow in the network, i.e. matrices $\cA$ and $\cB$, as well as on the specific performance measure, as will be detailed in Section~\ref{sec:EX}. In this article, we focus on the performance specifications expressed in the frequency domain, see~\cite{SkP:05}. For this purpose, let us further define the local, independent from the network, transfer function $T_i$ and the global transfer function of the network $T_{\bar w \rightarrow \bar z}$ between local ($r_i\rightarrow y_i$) and global ($\bar w \rightarrow \bar z$) signals respectively. 
Based on (\ref{module}), (\ref{excit1}) and (\ref{intercoM}) the following expression are obtained~:
\vspace*{-0.0cm}$$T_i(\sz,\theta_i) = \frac{K_i(\sz)G_i(\sz,\theta_i)}{1+K_i(\sz)G_i(\sz,\theta_i)} $$
$$ T_{\bar w \rightarrow \bar z}(\sz,\theta) = \dia_i(T_i (\sz,\theta_1)) \star \cM $$

The global performance specification will be deemed satisfactory if:
\vspace*{-0.2cm}\be  \label{eq:Glob_goal} 
\forall \omega,\ \bar \sigma\left( T_{\bar w \rightarrow \bar z}(\tom,\theta)\right) < W(\omega)
\vspace*{-0.2cm}\ee where $\tom$  defines $j\omega$ in the continuous time domain or $e^{j\omega}$ in the discrete time domain.

It is thus necessary to design (or redesign) the local controllers in order to ensure (or improve) the network performance and respect~(\ref{eq:Glob_goal}) with $\theta = \to$. 
However, since $\to$ is unknown, it will be necessary to identify a model for each of the systems $\cS_i (\tio)$. We assume that there is an identification procedure leading to a consistent parameter vector estimate $\thi$ of each subsystem true parameter vector $\tio$ as well as an estimate of the corresponding covariance matrices $\pti$. 
Such an identification procedure exists in open or closed-loop for each module independently, see~\cite{ljung1999system,BBHS:08}, or when the modules are connected to the network~\cite{BKHS:17}. 
 It implies with some probability that the true parameter vector $\tio$ belongs to some uncertainty set $U_i$ defined as :
\be \label{Ui} U_i=\{ \theta_i \ | \ (\theta_i-\thi)^T \pti^{-1} (\theta_i-\thi)<\chi\} \ee with a constant $\chi$ given the probability level we would like to ensure and the number of elements in the parameter vector~$ \to$.

 
 We also assume that there is a design procedure allowing to compute local controllers $K_i(\sz)$ such that the nominal global transfer function $T_{\bar w \rightarrow \bar z}(\sz,\theta)$, with $\theta =\th$ an estimate of $\to$, respects the frequency dependent bound~(\ref{eq:Glob_goal}). Such design procedures could be found in~\cite{ScD:01,KSCB:16}.

Of course since $\th$ is not necessarily equal to $\to$ this will not necessarily ensure the constraint~(\ref{eq:Glob_goal}) for the true system. In order to ensure the performance of the true system, in this article we would like to solve the following worst-case robustness analysis problem. Since $\tio \in U_i$ for all $i$, it is possible to ensure~(\ref{eq:Glob_goal}) with $\theta = \to$ by computing the worst-case gain of $T_{\bar w \rightarrow \bar z}(\tom,\theta)$, evaluated in terms of maximum singular values, $\forall\theta_i \in U_i$. Similarly to the robustness analysis approaches~\cite{Doy:82,Saf:82,FTD:91}, this computation will be performed frequency by frequency assuming an appropriate definition of the frequency gridding vector $\Omega = \{ \omega_1,\dots,\omega_{N_\omega}\}$ and that the properties ensured $\forall \omega_j \in \Omega$ imply that they are ensured $\forall \omega \in \reals$.

\begin{problem} \label{pbm:WCA}
	Given system~(\ref{module})-(\ref{interco}),(\ref{intercoM}), given 
	uncertainty sets~(\ref{Ui}), compute for each $\w_j\in\Omega$:
		\[ \min_{\theta_i\in U_i(i=1\dots \Nm)} \ \gamma(\w_j) \ \ \ \mbox{subject to}\]
	\be \bar \sigma\left( T_{\bar w \rightarrow \bar z}(\tom_j,\theta)\right) < \gamma(\w_j)  \label{eq:WCA_Problem} \ee 	
\end{problem}

If the minimal solution of the previous problem respects 
$$\gamma(\w_j)\leq W(\w_j)$$ for all $j$, then the computed controllers ensure that the true system $T_{\bar w \rightarrow \bar z}(\tom,\tho)$ respects the frequency dependent bound in~(\ref{eq:Glob_goal}) and thus the global performance.

Problem~\ref{pbm:WCA} is close to the well-known problem of worst-case robustness analysis (or $\mu$-analysis procedure) from the Robust Control Community~\cite{SkP:05}. However the uncertainty sets~(\ref{Ui}), representing ellipsoids in parameter space, are not the traditional ones considered in this field. The adaptation of traditional worst-case robust analysis methods to the case of the uncertainty set obtained from the identification can be found in~\cite{BGSA:01,SBBF:07,BBHS:08}. However direct application of these results in the case of a large-scale network system, i.e. when $\Nm$ is large, is not possible due to the high system complexity implying prohibitive computation time. As was mentioned in the introduction, the main contribution of this technical report is to extend these methods to the network context i.e. to derive tractable robustness performance analysis conditions while keeping computation time reasonable.


\section{Hierarchical Analysis Approach}

\subsection{Keys ideas}
As was discussed previously, the direct application of the worst-case analysis method will result in a prohibitive computation time for large scale networks. To avoid this, we propose to use the hierarchical robustness analysis approach of~\cite{DKS:13,DKS:14}. 

The main idea of the hierarchical approach is to decompose the network into two or more hierarchical levels and to perform the robustness analysis level by level by propagating the analysis results from one level to another. For some network systems such decomposition appears naturally, as for example for the system under consideration in this technical report : \emph{(i) local hierarchical level} : subsystem dynamics $T_i(\sz,\theta_i)$ defined by~(\ref{module}) and~(\ref{excit1}) and \emph{(ii) global hierarchical level}: the global information exchange~(\ref{interco}) and~(\ref{intercoM}). The robustness analysis at each hierarchical level allows to embed the \emph{subsystem dynamics} with a possibly complex non-linear dependence on the uncertainty, into a much simpler \emph{subsystem description} with a convex dependence on the uncertainty. We will call it the \emph{embedding procedure} in the sequel. Then in the next hierarchical level, the subsystem is replaced by this simple description and the procedure is repeated once again until reaching the last hierarchical level. The last step consists in the worst-case robustness analysis based on the propagated subsystem descriptions in order to evaluate the global network performance \emph{i.e.} solve the Problem~\ref{pbm:WCA}. The complexity and time computation reduction is ensured thanks to the embedding procedures and by the fact that all embeddings at each hierarchical level are independent and thus can be easily performed in parallel.

In this technical report, a two level hierarchical structure (local and global) is under consideration. Before formalizing this approach separately for the local and global hierarchical levels, let us first define what we mean by subsystem dynamics and subsystem description.

Since the performance measure in this technical report is expressed in the frequency domain, see~(\ref{eq:Glob_goal}), the subsystem dynamics are defined by the structured frequency response set $\cT^s_i(\w)$ of the subsystem transfer function  at frequency $\w$: 
\vspace*{-0.0cm}\be \label{eq:struct_set}\cT^s_i(\w) = \{ T_i(\tom,\theta_i)\ | \ \theta_i\in U_i\}\vspace*{-0.0cm}\ee 
The subsystem description in turn is defined by an uncertainty set $\cT_i(x_i(\w),y_i(\w),z_i(\w))$ of complex numbers 
 $\Delta_i(\w)\in \mathbf{C}$ 
 that respects a frequency dependent quadratic constraint imposed by $x^i(\w)\in \mathbf{R}$, $y^i(\w)\in \mathbf{C} $, $z^i(\w) \in \mathbf{R}$:
\vspace*{-0.0cm}\be \label{eq:UnstSet}
\begin{array}{c}
\cT_{i}(x^i(\w),y^i(\w),z^i(\w))=\left\{ \Delta_{i}(\w)\ \right|\\[1.5ex]
\left.\left[\begin{array}{c}
\Delta_{i}(\w)\\
1
\end{array}\right]^{*}\left[\begin{array}{cc}
x^i(\w)     & y^i(\w)\\
y^i(\w)^{*} & z^i(\w)
\end{array}\right]\left[\begin{array}{c}
\Delta_{i}(\w)\\
1
\end{array}\right]\leq0\right\} 
\end{array}
\ee
Let us introduce the following definition characterizing the frequency response of a system :
\begin{definition}[Dissipativity] \label{def:diss}
\emph{An LTI system $H(\sz)$ is $\left\lbrace x(\w),y(\w),z(\w)\right\rbrace$ - dissipative at $\omega$ for some $x(\w)\in \mathbf{R}$, $y(\w)\in \mathbf{C} $, $z(\w) \in \mathbf{R}$, if its frequency response $H(\tom)$ respects the following quadratic constraint at $\omega$:
	$$
	\left[\begin{array}{c}
	H(\tom)\\
	1
	\end{array}\right]^{*}\left[\begin{array}{cc}
	x(\w)     & y(\w)\\
	y(\w)^{*} & z(\w)
	\end{array}\right]\left[\begin{array}{c}
	H(\tom)\\
	1
	\end{array}\right]\leq0.
	$$   }
\end{definition}
 
If the following additional constraint is imposed on $x(\w)$, then the corresponding quadratic constraint defines a convex set :
\vspace*{-0.2cm}\be \label{eq:xyz_convx} x(\w)\geq 0.\vspace*{-0.2cm}\ee
Please note that, in the case of $x(\w)>0$, by Definition~\ref{def:diss} and the Schur complement \cite{BEFB94}, the following constraint is implied: $y^2(\w) \geq x(\w)z(\w)$.  When $x(\w)=0$, no constraint is imposed on $y(\w)$ and $z(\w)$. In order to reduce the computational complexity, the convexity constraint~(\ref{eq:xyz_convx}) will be used in the sequel.

If each subsystem $T_i(\sz,\theta_i)$ is $\left\lbrace x_i(\w),y_i(\w),z_i(\w)\right\rbrace$ - dissipative for some frequency dependent $x_i(\w),y_i(\w),z_i(\w)$ and for all $\theta_i \in U_i$ and $\forall \omega$, we then obtain the following embedding $\cT^s_i(\w) \subset \cT_i((x_i(\w),y_i(\w),z_i(\w)),\ \ \forall \omega$; and the frequency responses of the uncertain subsystems $T_1(\sz,\theta_i),\dots,T_{N_{mod}}(\sz,\theta_i)$ generated by varying $\theta_i \in U_i$, can be replaced in the global hierarchical level by the corresponding subsystem description $\cT_i((x_i(\w),y_i(\w),z_i(\w))$.

Of course, since the set $\cT_i((x_i(\w),y_i(\w),z_i(\w))$ is in general larger than the set $\cT^s_i(\w)$ the result of the corresponding worst-case analysis might be conservative. In order to reduce this conservatism, it is important to choose suitable $x_i(\w),y_i(\w),z_i(\w)$ for each subsystem defining as tight embedding as possible. It is also possible to compute several complementary triplets $x^k_i(\w),y^k_i(\w),z^k_i(\w)$ for $k=1\dots N_d$ defining therefore $N_d$ dissipativity properties for each subsystem. It allows to define for each subsystem a basis of dissipativity properties (a set of subsystem descriptions) and propagate it to the global hierarchical level. 
Such a suitable choice in the context of the uncertainty set~(\ref{Ui}) obtained through an identification procedure is presented in the next subsection while Subsection~\ref{subsec:Global_step} presents how the embeddings are combined and propagated in a global hierarchical step in order 
to efficiently solve Problem~\ref{pbm:WCA}. It is clear that the more dissipativity characterizations are used for each subsystem, the more the conservatism is reduced. Of course, the price to pay for this is the increase of computation time. For this reason it is important to find appropriate triples $x^k_i(\w),y^k_i(\w),z^k_i(\w)$ at each hierarchical step.

\subsection{Local Step} \label{subsec:Local_step}

In this subsection we present how to efficiently compute different dissipativity triplets $x ,y ,z$ at a given frequency $\w$ such that an uncertain system $T(\sz,\theta_i)$ is \diss, $\forall \theta_i \in U_i$ with $U_i$ defined in~(\ref{Ui}).

For this purpose let us define the following factorization of the transfer function $T(\sz,\theta_i)$, suitable for the system identification~\cite{BGSA:01}:
\vspace*{-0.2cm}\be \label{eq:T_fact}
T(\sz,\theta_i) = \frac{e(\sz) + Z_N(\sz) \theta_i}{1 + Z_D(\sz) \theta_i}
\ee with $\theta \in \mathbf{R}^{n_\theta}$ and then present the following Lemma.

\begin{lemma}
	\label{lem:Diss}
	Given the uncertain LTI system $T(\sz,\theta_i)$ in~(\ref{eq:T_fact}), it is \diss  
	\ for all $\theta_i\in U_i$ and for given $\w$, $x(\w)\in \mathbf{R}$, $y(\w)\in \mathbf{C} $, $z(\w) \in \mathbf{R}$ respecting~(\ref{eq:xyz_convx}), if and only if \\
	$(i)$ in the case of $x(\w)>0 \ :$
	\be \bmtr{c|c} -\alpha(\w)  & \lambda(\w) \\  \hline \lambda^*(\w) & -A_1(\w) - \xi(\w) B + j \cX(\w)  \emtr\leq 0 \label{embeddingxyz1} \ee 
	$(ii)$ in the case of  $x(\w)=0 \ :$	
	\be  
	A_2^*(\w)y(\w)+y^*(\w)A_2(\w) + A_1(\w)z(\w)-\xi(\w)B+j \cX(\w)  \leq 0 \label{embeddingxyz2} \ee
	with {\small  $\lambda(\w)=\bmtr{c|c} Z_{N}(\tom)+\frac{y(\w)}{x(\w)}Z_{D}(\tom) & e(\tom)+\frac{y(\w)}{x(\w)}\emtr$, 
	 $$ A_1(\w) = \bmtr{cc} Z^*_{D}(\tom)  Z_{D}(\tom)  & Z^*_{D}(\tom) \\ Z_{D}(\tom) & 1 \emtr,$$
	  $$ A_2(\w) = \bmtr{cc} Z^*_{D}(\tom)  Z_{N}(\tom)  & Z^*_{D}(\tom)e(\tom) \\ Z_{N}(\tom) & e(\tom) \emtr,$$
	  $\alpha(\w) =  \frac{y^{2}(\w)}{x^{2}(\w)} - \frac{z(\w)}{x(\w)}$, $\ B = \bmtr{cc} \pti^{-1} & -\pti^{-1}\th_i \\ -\th_i^T \pti^{-1} & \th_i^T \pti^{-1} \th_i-\chi \emtr $} and \\[1ex]
  \ni some $\xi(\w)\geq 0\in \mathbf{R}$, $\cX(\w)=-\cX^T(\w)\in \mathbf{R}^{n_\theta\times n_\theta}.$ \\
\end{lemma}
\begin{proof} For the sake of conciseness, we will drop the frequency argument $\w$ and $\tom$ in the variables. By definition of dissipativity, $T(\sz,\theta)$ is $\{x,y,z\}$ - dissipative $\forall \theta \in U$, is equivalent to :
\vspace*{-0.2cm}	$$ (i) \left(T(\theta_i) + \frac{y}{x}\right)^*\left(T(\theta_i) + \frac{y}{x}\right) \leq  \frac{y^{2}}{x^{2}} - \frac{z}{x}, \text{for }  x>0 $$
	
	\vspace*{-0.4cm}	$$ (ii) \left(y^*T(\theta_i) \right)^* + y^*T(\theta_i) + z \leq 0, \text{for } x=0, $$ $\forall \theta_i \in U_i$.	Using factorization~(\ref{eq:T_fact}), and compact notation \newline $\bar\theta=[\theta_i^T \ 1]^T$, the previous inequalities are equivalent to 
\vspace*{-0.2cm}	\be \label{equ_xyz_th1}(i) \  \bar \theta^T \left (-A_1 - \lambda^* \frac{-1}{\alpha} \lambda \right ) \bar \theta \leq  0, \ \ \forall \theta_i \in U_i \vspace*{-0.2cm}\ee
\vspace*{-0.2cm}\be \label{equ_xyz_th2}(ii) \  \bar \theta^T \left (A_2^*y+y^*A_2  + A_1 z\right ) \bar \theta \leq  0, \ \ \forall \theta_i \in U_i \ee
	
	\ni while the constraint $\theta_i \in U_i$ is equivalent to $\bar \theta^T B \bar \theta<0$. Consequently, by virtue of the S-procedure \cite{BEFB94} and Lemma 2 in \cite{BSGHH:05},~(\ref{equ_xyz_th1}) or~(\ref{equ_xyz_th2}) holds if and only if there exist $\xi\geq 0$ and $\cX=-\cX^T$ such that 
	\vspace*{-0.2cm}	\be \label{equ_xyz_th1_S}(i) \ - A_1 - \lambda^* \frac{-1}{\alpha} \lambda - \xi B + j \cX \leq 0 \vspace*{-0.2cm}\ee 
	\vspace*{-0.2cm}\be \label{equ_xyz_th2_S}(ii) \ A_2^*y+y^*A_2 + A_1 z - \xi B + j \cX \leq  0 \ee 
	The last constraint is exactly condition~(\ref{embeddingxyz2}). Due to convexity constraint~(\ref{eq:xyz_convx}), with non zero $x\neq0$,  
	$\alpha>0$, and the application of the Schur complement \cite{BEFB94} shows that~(\ref{equ_xyz_th1_S}) is equivalent to~(\ref{embeddingxyz1}). This concludes the proof. 
\end{proof}

Please note that the sufficiency of Lemma~\ref{lem:Diss} can be proved using the result of~\cite{DKS:14} (see Corollary 2.2). As is shown in the proof of Lemma~\ref{lem:Diss}, the result of~\cite{DKS:14} is adapted to the case of uncertain vectors that belong to an ellipsoid which recovers sufficient and necessary conditions of $\left\lbrace x(\w),y(\w),z(\w)\right\rbrace$ - dissipativity. This lemma is an extension of the robustness analysis result of~\cite{BKHS:17} and will be used to generate different types of embeddings.

We will now consider two types of embedding: the disc and the band embedding, and formulate a convex optimization problem to compute them. Please note that thanks to Lemma~\ref{lem:Diss}, it is possible to study other types of embedding, as for example cone embedding~\cite{LKSM:15}, half planes etc.
\subsubsection{Disc Embedding}
 Given system $T(\sz,\theta)$ in~(\ref{eq:T_fact}), its frequency response set $\{ T(\tom,\theta_i)\ | \ \theta_i\in U_i\}$ is embedded in a disc set at $\w$ if 
 \vspace*{-0.2cm}\be | T(\tom,\theta_i) - c(\w) |\leq \rho(\w), \ \ \ \forall \theta_i \in U_i \label{eq:embeddin_disc} \ee where $c(\w)\in \mathbf{C}$ is the center of the disc and $\rho(\w)\in \mathbf{R}$ is its radius, see~\cite{DKS:14}. 
 The size measure of this embedding is the radius of the disc, and the problem of the computation of the tightest embedding can be formulated as follows assuming appropriate gridding $\Omega$.
 
\begin{problem} \label{pbm:Disk_emb}
	Given system~(\ref{eq:T_fact}) and its uncertainty set~(\ref{Ui}), compute for each $\w_j\in\Omega$:
	\vspace*{-0.2cm}	\[ \min_{\rho(\w_j), c(\w_j)} \ \rho(\w_j) \ \ \ \mbox{subject to~(\ref{eq:embeddin_disc}) with $\omega=\omega_j$}\]
\end{problem}
This problem is efficiently solved by the following theorem.

\begin{theorem}[Disc embedding]\label{thm:Disk_emb}
	Given system~(\ref{eq:T_fact}) and its uncertainty set~(\ref{Ui}), Problem~\ref{pbm:Disk_emb} is solved by the following convex optimization problem:
\vspace*{-0.2cm}	\be \label{eq:LMI_disc_emb}  \min_{\rho^2(\w_j), c(\w_j)} \ \rho^2(\w_j) \mbox{ s.t.~(\ref{embeddingxyz1}) is holds with $\omega=\omega_j$ and} \vspace*{-0.0cm}\ee
	$$x(\w_j) = 1,\  y(\w_j) = -c(\w_j),\ z(\w_j) = c^2(\w_j) - \rho^2(\w_j).$$
\end{theorem}
	\begin{proof}
	The roof is straightforward after replacing the value of $x,y,z$ and applying Lemma~\ref{lem:Diss}. Please note that in this case $\alpha(\w_j) = \rho^2(\w_j)>0$ and 
	$\lambda(\w_j) = \bmtr{c|c} Z_{N}(\tom_j)-c(\w_j)Z_{D}(\tom_j) &  e(\tom_j)-c(\w_j) \emtr $, implying affine dependence on the decision variables. As a consequence, the optimization~(\ref{eq:LMI_disc_emb}) is an LMI optimization and can be solved efficiently.
	\end{proof} 

\subsubsection{Band Embedding}
Given system $T(\sz,\theta_i)$ in~(\ref{eq:T_fact}), its frequency response set $\{ T(\tom,\theta_i)\ | \ \theta_i\in U_i\}$ is embedded in a band set at $\w$ if $\forall \theta_i \in U_i$
\vspace*{-0.1cm}\be 2a_2(\w)\leq T^*(\tom,\theta_i)n(\w) + n^*(\w)T(\tom,\theta_i) \leq 2a_1(\w), \label{eq:embeddin_band} \vspace*{-0.1cm}\ee where $n(\w)\in \mathbf{C}$ is the complex number which defines the vector $\overrightarrow{n} = [Re(n),Im(n)]^T$ giving the band orientation in complex plain (it is perpendicular to both band hyperplanes) and $a_1(\w),a_2(\w)\in \mathbf{R}$ are the signed distances of the two band hyperplanes to the origin multiplied by $|n|$, see~\cite{DKS:14}. The size measure of this embedding is the band width $d(\w) = a_1(\w)-a_2(\w)$ (see~\cite{DKS:14} and Fig.~\ref{fig:Loc_step} for illustration),
 and the problem of computation of the tightest embedding can be formulated as follows assuming again appropriate gridding $\Omega$.

\begin{problem} \label{pbm:Band_emb}
	Given system~(\ref{eq:T_fact}) and its uncertainty sets~(\ref{Ui}), compute for each $\w_j\in\Omega$:
	\vspace*{-0.1cm}	\[ \min_{n(\w_j),a_1(\w_j),a_2(\w_j)} a_1(\w_j)-a_2(\w_j) \ \ \mbox{subj. to~(\ref{eq:embeddin_band}) with $\omega=\omega_j$}\]
\end{problem}
This problem is efficiently solved by the following Theorem.

\begin{theorem}[Band embedding]
	Given system~(\ref{eq:T_fact}) and its uncertainty sets~(\ref{Ui}), Problem~\ref{pbm:Band_emb} is solved by the following convex optimization problem:
\vspace*{-0.2cm}	\be \label{eq:LMI_band_emb}  \min_{a_1(\w_j),a_2(\w_j), n(\w_j)} \  a_1(\w_j)-a_2(\w_j)\ee	
	s.t.~(\ref{embeddingxyz2}) holds with $\omega=\omega_j$ and
	$$x_1(\w_j) = 0,\  y_1(\w_j) = n(\w_j),\ z_1(\w_j) = -2a_1(\w_j)$$ 
	and~(\ref{embeddingxyz2}) holds with $\omega=\omega_j$ and 
	$$x_2(\w_j) = 0,\  y_2(\w_j) = -n(\w_j),\ z_2(\w_j) = 2a_2(\w_j).$$
\end{theorem}
	\begin{proof}
		The proof is straightforward after replacing the value of $x_1,y_1,z_1$, $x_2,y_2,z_2$ and applying Lemma~\ref{lem:Diss} twice. Please note that, in this case as well, the dependence on the decision variables is affine. As a consequence, the optimization~(\ref{eq:LMI_band_emb}) is an LMI optimization and can be solved efficiently.
	\end{proof}

\subsection{Global Step} \label{subsec:Global_step}
In this subsection, we assume that for all $\w_j \in \Omega$ and for each subsystem $T_i(\sz,\theta_i)$, several embeddings are found in the local step. We thus obtain $N_d$ dissipativity triplets $x^k_i(\w_j),y^k_i(\w_j),z^k_i(\w_j)$ for $k=1,\dots,N_{d}$, for each subsystem $j=1,\dots,N_{mod}$ and for all $\w_j\in \Omega$.
The next theorem allows to compute an upper bound $ \gamma_{UB}(\w_j)$ on the maximum amplification $\gamma(\w_j)$ of Problem~\ref{pbm:WCA}.

\begin{theorem}
Given system~(\ref{module})-(\ref{interco}),(\ref{intercoM}), a frequency $\w_j$ and  given $x^k_i(\w_j),y^k_i(\w_j),z^k_i(\w_j)$ such that 
$$\cT^s_i(\w_j) \subset \cT_i((x^k_i(\w_j),y^k_i(\w_j),z^k_i(\w_j))$$
for $k=1,\dots,N_{d}$, $i=1,\dots,N_{mod}$ (see~(\ref{eq:struct_set}) and~(\ref{eq:UnstSet}))

The upper bound $ \gamma_{UB}(\w_j)$ on the maximum amplification $\gamma(\w_j)$ of Problem~\ref{pbm:WCA} is the solution of the following LMI optimization problem:

\vspace*{-0.2cm}{\small \[ \min_{\bar\gamma^2(\w_j),T^k_{\omega},(k=1\dots N_d)} \ \ \ \bar \gamma^2(\w_j)\] \be \label{sepgr} s.t. \ \bvec{c} M(\tom) \\ I \evec^* \cN(\bar\gamma^2(\w_j))  \bvec{c} M(\tom) \\ I \evec > 0,  \mbox{ with}\ee

		$$ 
	\cN(\bar\gamma^2(\w_j))\eqbydef  \bmtr{c|c}  \begin{array}{c|c} \sum T^k_\w \cZ^k_d & 0\\ \hline 0 & -I  \end{array} &  \left[ \begin{array}{c|c} \sum T^k_\w {\mathcal Y^k_d} & 0\\ \hline 0 & 0  \end{array}\right]^*  \\  \hline  \begin{array}{c|c} \sum T^k_\w {\mathcal Y^k_d} & 0\\ \hline 0 & 0  \end{array} &  \begin{array}{c|c} \sum T^k_\w \cX^k_d & 0\\ \hline 0 & \bar \gamma^2(\w_j)I  \end{array} \emtr$$


} 
	
	\ni with strictly definite positive diagonal matrices $T^k_\w\in \reals^{\Nm \times \Nm},$ and $k=1\dots N_d$.
{\small	$$\cX^k_d = \dia_i(x_{i}^{k}(\w_{j})), \ 
	\cY^k_d = \dia_i(y_{i}^{k}(\w_{j})), \
	\cZ^k_d = \dia_i(z_{i}^{k}(\w_{j}))$$}
\end{theorem}

\begin{proof}
This theorem can be straightforwardly deduced from the separation of graph theorem~\cite{Saf:80} and from the results in~\cite{DKS:14}. It follows from the fact that the constraint~(\ref{sepgr}) is a sufficient condition for $\bar\sigma \left( \cT(\omega_j) \star \cM (\omega_j)\right)  <\gamma_{UB}^2(\w_j)$, with \vspace{-0.0cm}
$$\cT(\omega_j) = \dia_i\left( \Delta_i (\omega_j)\right),$$
$ \forall \Delta_i(\omega_j) \in  \bigcap_k\left\lbrace    \cT_i(x^k(\w_j),y^k(\w_j),z^k(\w_j))\right\rbrace $ to hold. \vspace{-0.45cm}

\end{proof} 

\section{Computational Complexity}
As already mentioned earlier, Problem~\ref{pbm:WCA} can be solved directly by the the method proposed in~\cite{SBBF:07,BBHS:08}. Let us call this approach the \emph{direct worst case analysis approach}. In fact, in the case of a SISO global transfer function $T_{\bar w \rightarrow \bar z}(\sz,\theta)$, Lemma~\ref{lem:Diss} together with Theorem~\ref{thm:Disk_emb} can be seen as a generalization of the result~\cite{SBBF:07,BBHS:08}. Indeed, considering factorization of $T_{\bar w \rightarrow \bar z}(\sz,\theta)$ similar to~(\ref{eq:T_fact}), defining overall parameter vector $\theta$ and its uncertainty similar to~(\ref{Ui}), Problem~\ref{pbm:WCA} is equivalent to Problem~\ref{pbm:Disk_emb} with center $c(\omega_j)=0$ and $\gamma(\omega_j) = \rho(\omega_j), \ \forall j$. It can thus be efficiently solved by convex optimization in Theorem~\ref{thm:Disk_emb}. This result can be generalized to the case of Multi-Input, Multi-Output (MIMO) global transfer function $T_{\bar w \rightarrow \bar z}(\sz,\theta)$. However the necessity part of the result will be lost so let us focus here on the SISO case only. 

In this technical report, the hierarchical worst case analysis approach is proposed. The main interest of the hierarchical approach is the computational time reduction in comparison to the direct one and in the case of large scale network $\Nm\gg 1$. To evaluate this reduction in both cases, independently of the computational facilities, let us assume that the computational time is equal to the algorithm complexity and let us investigate its evolution as a function of the subsystem number $\Nm$. For interior point methods for LMI optimization, it is a cubic function of the decision variable number $n$~\cite{NeN:93}:
$$ t =\mathcal O (n^m) \eqbydef \alpha_mn^m + \dots + \alpha_1n + \alpha_0  $$ with some non-negative $\alpha_i$ and $m=3$.
 
Supposing that each subsystem is SISO with the same size $\forall i,\ n_{\theta_i} = \bar n_\theta$ of parameter vector $\tio$, the amount of decision variables of the direct worst case analysis approach is equal to 
$$2+ \frac{n_\theta (n_\theta-1)}{2} = 2+ \frac{N_{mod}\bar n_\theta (\Nm \bar n_\theta-1)}{2}$$ which gives polynomial dependence of the order $6$ for the direct approach time computation $t^{direct}$ with respect to the number of subsystems $N_{mod}$ :
$$ t^{direct} =\mathcal O (N_{mod}^6). $$

For the hierarchical approach, with two hierarchical levels, we obtain $N_{mod}n_d$ local embeddings with $3 + \frac{\bar n_\theta (\bar n_\theta-1)}{2}$ decision variables each and one global analysis with $N_{mod}n_d$ decision variables. It gives the overall computation time~: 
$$ t^{hierarch.} = N_{mod}n_d \mathcal O (\bar n_\theta^6) + \mathcal O (N_{mod}^3) \approx \mathcal O (N_{mod}^3) $$ and in the case of parallel computation of local embeddings : 
$$ t^{hierarch.}_{parallel} = n_d \mathcal O (\bar n_\theta^6) + \mathcal O (N_{mod}^3) \approx \mathcal O (N_{mod}^3).$$
Therefore, if the parallel computation of the local subsystem embeddings is allowed by available computational facilities, the time computation reduction is even better.
As a consequence, the hierarchical approach for $N_{mod}\gg 1$ is much more efficient from computational point of view while, as illustrated in the next section, keeping reasonable conservatism level with an appropriate choice of embeddings.

\section{Numerical Example} \label{sec:EX}
Let us now consider an illustration example of an Automated Highway System (AHS): a platoon of autonomous cars following external reference signals as in~\cite{SPH:04}. Each car's simplified model dynamics is described by~(\ref{module}), with $G_i(s,\theta) = \frac{k_i}{s^2(\tau_is + 1)}$ and true parameter vector $\tio = [\tau_i,k_i]^T$ where $\tau_i,k_i$ were randomly chosen around $0.105$ and $0.95$ respectively with uniform $\pm 10\%$ distribution. Each system is controlled by the same initial decentralized controller $K_{init}(s) = \frac{2s + 1}{0.05s + 1}$ taken from~\cite{SPH:04}, see~(\ref{excit1}). There are $N_{mod}=5$ cars in the network which are allowed to exchange information according to bidirectional chain topology, see~\cite{SPH:04}, as depicted in~Fig.\ref{figexnet} and defined by~(\ref{excit1}). 

The main objective of the network is that each car follows a ramp reference signal $ref(t)$, available only for the first car, shifted by a constant value $\delta_i = i\delta,\ \forall i$, while keeping string instability (oscillation propagation through the network) limited~\cite{SPH:04}. It can be shown that this tracking performance specification is equivalent to the ability of each car to track the same ramp signal $ref(t)$ ensuring that all local tracking errors $e_i = r_i - y_i$ go to zero in steady-state. It is sufficient to locally apply a constant shift $-\delta \neq 0$ at the input of each subsystem, as depicted in Fig.~\ref{figexnet}, and to perform a suitable change of variable $y_i \rightarrow y_i-\delta$.

\begin{figure}[thpb]
	\centering  \hspace{-0.5cm}
\parbox{3in}{ \vspace{-0.5cm} \includegraphics[width=250pt]{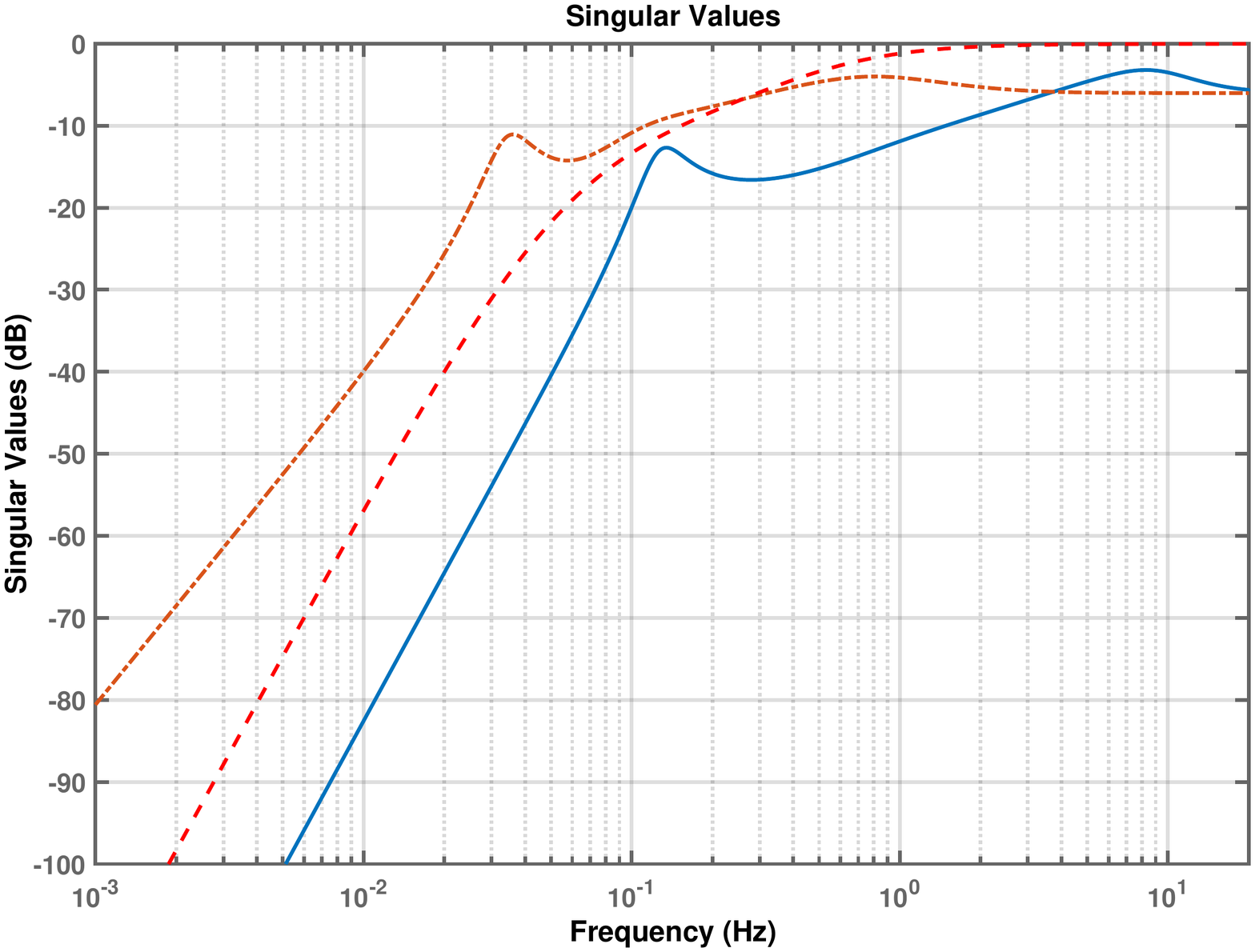}	}
	\caption{Maximal singular value of the true system $T_{\bar w \rightarrow \bar z}(\sz,\theta_o)$ for initial controller (orange dash-dotted line), improved controller (blue solid line) and imposed frequency constraint $W(\omega)$ (red dashed line).\vspace{-0.2cm}}
	\label{fig:SV_NET}
\end{figure}

As a consequence, let us define performance input $\bar w(t) = ref(t)$ and performance output $\bar z(t) = \bar r(t) - \bar y(t)$. It thus determines the interconnection topology~(\ref{intercoM}) with $\mathcal{M} = \left[ \begin{array}{cc}
\mathcal{A} & \mathcal{B} \\
\mathcal{A}-I & \mathcal{B} 
\end{array}\right].$ If the maximum singular value of $T_{\bar w \rightarrow \bar z}(\sz,\theta_o)$ has a slope of $+40$ dB/dec at low frequency range, then the tracking performance is ensured, see~\cite{SkP:05}. Moreover, a lower gain ensures a better tracking speed and the resonance peak limitation reduces the effects of string instability~\cite{SPH:04}. 
The maximal singular value of the true system $T_{\bar w \rightarrow \bar z}(\sz,\theta_o)$ with initial controller $K_{init}$ is represented by orange dash-dotted line in Fig.~\ref{fig:SV_NET}. In order to improve the tracking performance of the network and to reduce the oscillation effects provoked by the string instability, let us impose the frequency constraint~(\ref{eq:Glob_goal}) with $W(\omega)$ represented in Fig.~\ref{fig:SV_NET} by the red dashed line.

\begin{figure}[thpb]
	\centering
	\parbox{3in}{ \vspace{-0.5cm} \includegraphics[width=250pt]{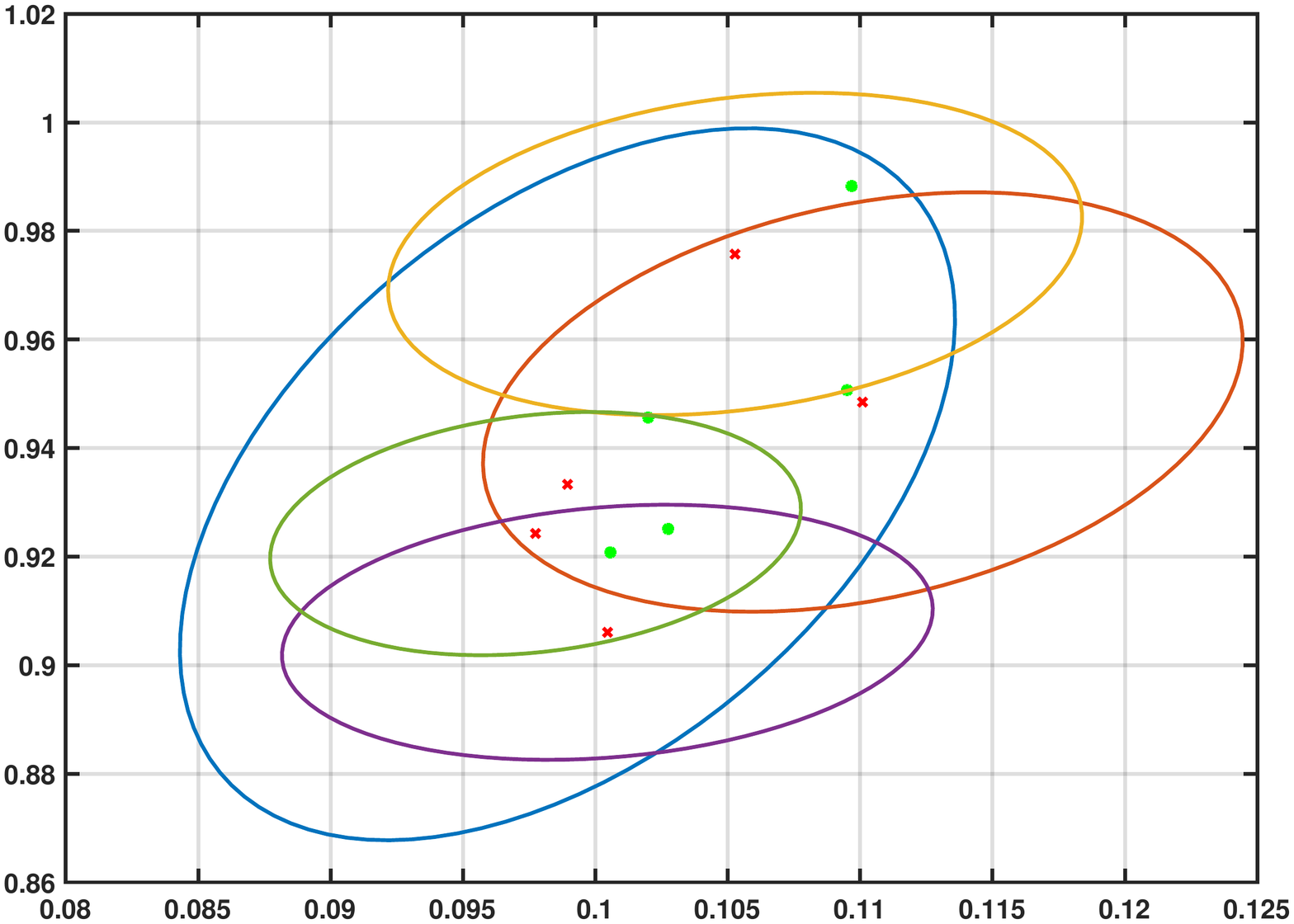}	}
	\caption{Identification results. True parameter vectors $\tio$ (green dots), its estimated values $\thi$ (red crosses), and corresponding ellipsoid set borders (full lines) for $\chi$ chosen to ensure $95\%$ probability.\vspace{-0.3cm}}
	\label{fig:Id}
\end{figure}

To satisfy this constraint, first an identification procedure is performed leading to a consistent parameter vector estimate $\thi$ of each subsystem true parameter vector $\tio$ as well as an estimation of the corresponding covariance matrices $\pti$ ensuring~(\ref{Ui}). Due to the presence of a double integrator in the car transfer function model, this identification experiment has to be performed in closed loop with a stabilizing controller either independently for each module~(see~\cite{ljung1999system,BBHS:08}) or in the network~(see~\cite{BKHS:17}). The results of the latter method are presented in Fig.~\ref{fig:Id} where the controllers were chosen as $K_i(s) = K_{init}(s),\ \forall i$. Different discrete-time white noise excitation signals of length $N_{id}=1000$, sampling time $T_s=0.01$ sec and variance $10$ are added via a zero order hold to the references $r_i$ of each closed-loop systems $T_i(\sz,\tio)$. The measured discrete signal $y_i$ is also perturbed by
generated mutually independent white noise discrete signals $v_i$ with variance of $4$ each modeling the measurement noise effects. A standard, prediction-error identification criterion is used, see~\cite{ljung1999system}. Notice that in this example the continuous transfer function parameters $k_i$ and $\tau_i$ and the corresponding covariance matrices could be directly identified since the car transfer function model is rather simple.  
An adapted to the subsystem dynamics optimization, taking into account zero-order hold effects, had to be applied in order to identify directly continuous transfer function parameters $k_i$ and $\tau_i$ and  corresponding covariance matrices. 

A new improved decentralized controller is designed based on the $H_\infty$ framework~\cite{ScD:01,KSCB:16}: 
$$K(s) = \frac{12111(s+10)(s^2 + 0.9s + 0.4)}{s(s^2 + 111.6s + 6230)}.$$
It ensures that the nominal global transfer function $T_{\bar w \rightarrow \bar z}(\sz,\theta)$, with $\theta =\th$, respects the frequency dependent bound in~(\ref{eq:Glob_goal}), see Fig.~\ref{fig:SV_NET}.

\begin{figure}[thpb]
	\centering
	\parbox{3in}{ \vspace{-0.5cm} \includegraphics[width=250pt]{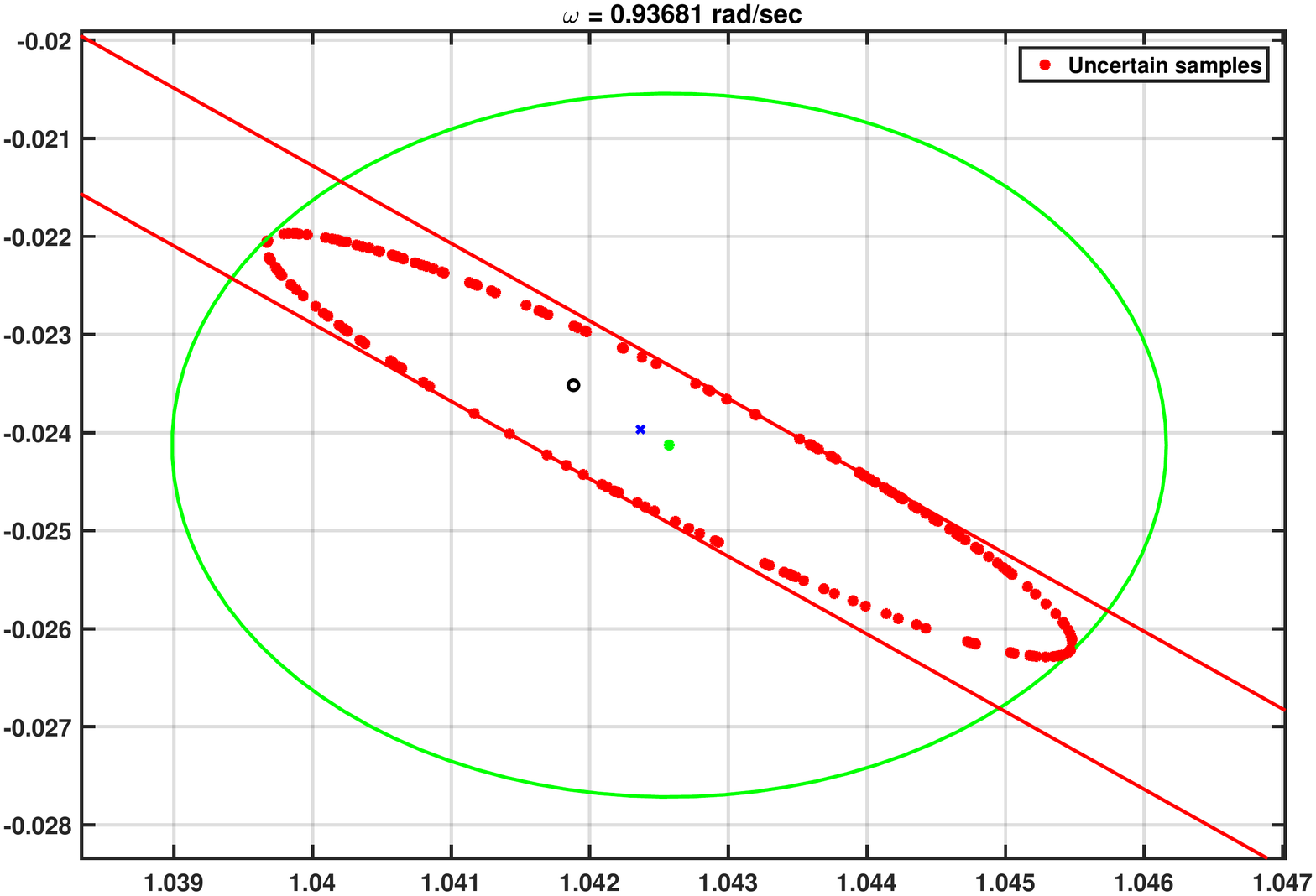}	}
	\caption{Local step embedding results. Borders of structured uncertainty set $\mathcal T^s_1$ (red dots), of the minimum radius disc embedding (green full circle), of the tightest band (red full lines), circle center (green dot), estimated frequency response (blue cross) and true frequency response (black round).}
	\label{fig:Loc_step}
\end{figure}

\begin{figure}[thpb]
	\centering 
	\hspace{-8mm}\parbox{3in}{ \vspace{-0.0cm} \includegraphics[width=250pt]{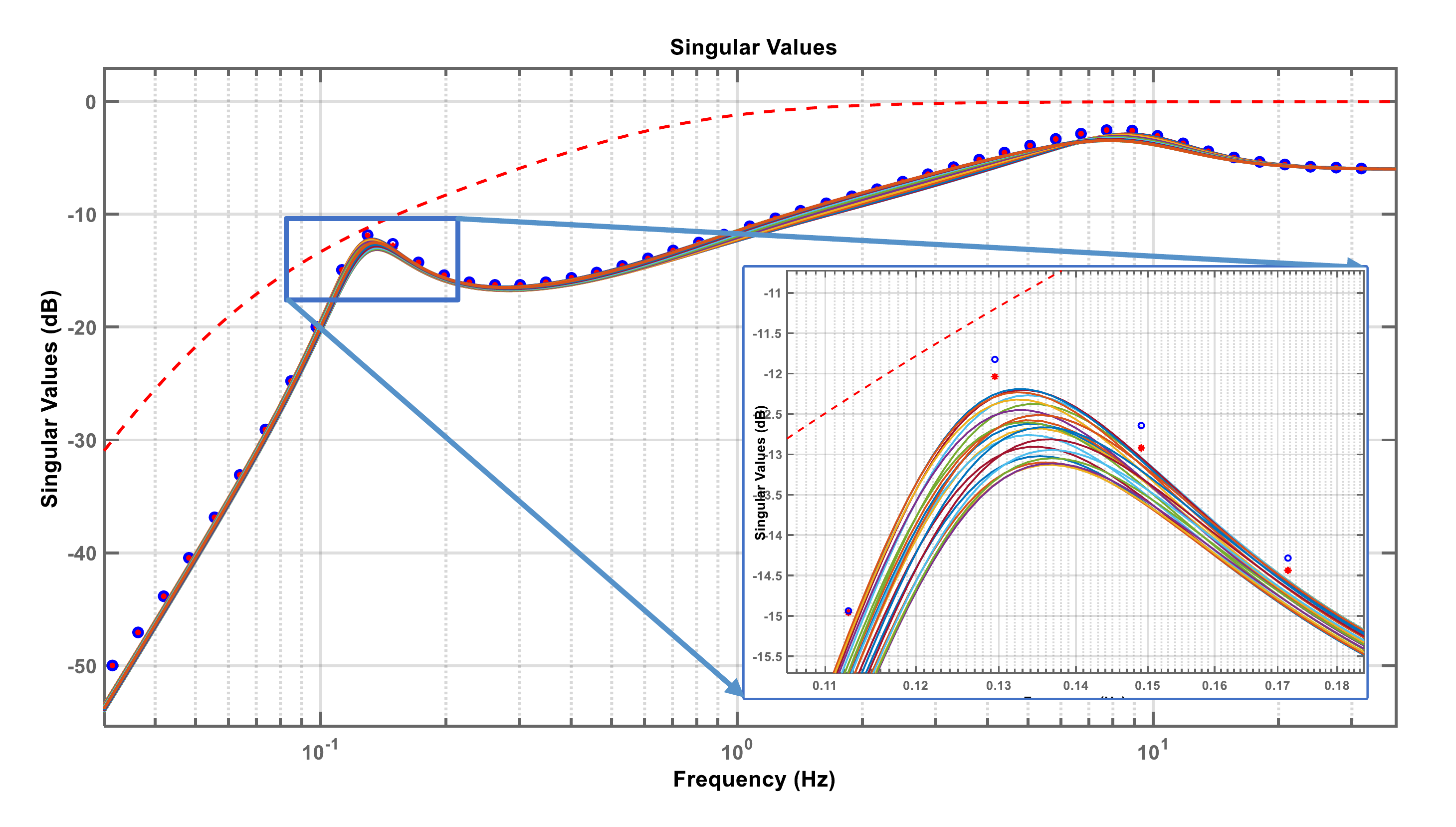}	}
	\caption{Global step analysis results. Upper bounds computed by \emph{(i)} propagation of disc embedding only (blue rounds), by \emph{(ii)} propagation of disc and band embeddings (red dots), Monte-Carlo samples of maximal singular value of $T_{\bar w \rightarrow \bar z}(\sz,\theta)$, for some $\theta_i\in U_i$.}
		\label{fig:Global_step}
	\end{figure}

Our problem is now to efficiently test if the constraint is satisfied by the true system by solving Problem~\ref{pbm:WCA} for properly chosen $\Omega$. To do so, the proposed hierarchical approach is used. The results of the local step embeddings for the first system and at $0.15$ Hz are presented in Fig.~\ref{fig:Loc_step} where the borders of the minimum radius disc embedding (green full circle) and of the tightest band (red full lines) are presented. For the sake of illustration reason, we show the borders of the structured uncertainty set $\mathcal T^s_1$ (red dots), the estimated $T_1(\th_1)$ (blue cross) and the true $T_1(\theta_{1,0})$ (black round) value of the corresponding frequency responses evaluated at $\omega = 0.15$ Hz. Notice that disk center $c(\w)\neq T_1(\jw,\theta_{1,0})$.  The results are found by solving the LMI optimization problems~(\ref{eq:LMI_disc_emb}) and~(\ref{eq:LMI_band_emb}). Similar results are obtained for other subsystems and other frequencies from $\Omega$. The global step analysis results are presented in Fig.~\ref{fig:Global_step} for two cases : computed $\gamma_{UB}$ based on the propagation of \emph{(i)} disc embedding only (blue rounds) and of \emph{(ii)} disc and band embeddings (red dots). Fig.~\ref{fig:Global_step} also presents some Monte-Carlo samples i.e. the maximal singular value of $T_{\bar w \rightarrow \bar z}(\sz,\theta)$ for randomly chosen $\theta_i\in U_i$. As we can see, the worst-case bounds are respected. Surprisingly even though the disc embedding set is much bigger than the intersection of disc and band sets (see Fig.~\ref{fig:Loc_step}), the overall upper bound $\gamma_{UB}$ is not improved a lot, see Table~\ref{tab:WCA_results}. It is due to the fact that, in this application, the phase uncertainty information, mostly captured by the band embedding, is much less important than the gain uncertainty information, mostly captured by the disc embedding. The corresponding computation times are also given Table~\ref{tab:WCA_results} for both serial and parallel computation of local embeddings. Finally, maximal singular values of the true system  $T_{\bar w \rightarrow \bar z}(\sz,\theta_o)$ with the new controller are represented by the blue solid line in Fig.~\ref{fig:SV_NET}.

\vspace{-0.1cm}
\begin{table}[h]
\caption{Hierarchical worst-case analysis results}
\label{tab:WCA_results}
\vspace{-0.4cm}
\begin{center}
\begin{tabular}{|c||c|c|c|}
\hline
 &  disc only  &  disc + band   & difference\\
\hline
\hline
$\gamma_{UB}$ @ $0.13$ Hz & $-11.83$ dB  & $-12.04$ dB &$1.8$\%\\
$\gamma_{UB}$ @ $0.15$ Hz & $-12.64$ dB  & $-12.92$ dB &$2.2$\%\\
$\gamma_{UB}$ @ $0.17$ Hz & $-14.28$ dB  & $-14.44$ dB &$1.1$\%\\
Overall Time & $15.33$ sec  & $19.04$ sec   &$-24.2$\%\\
Overall Time (Parallel) & $11.85$ sec  & $14.43$ sec & $-21.8$\%\\
\hline
\end{tabular}
\end{center}
\end{table}

\vspace{-0.05cm}

\section{Conclusions}

In this technical report we proposed robustness analysis method adapted to the uncertainty sets constructed by identification in a network context. The type of network in this system is usual in the literature of multi-agent systems and the size of the network plays a crucial role in the robustness analysis complexity.
In order to manage the trade-off between the computation time and the precision of the obtained result, the hierarchical robustness analysis approach was proposed and illustrated in the case of SISO subsystems. Future extension is the MIMO subsystem case with an appropriate choice of hierarchical structure (with possibly more than two hierarchical levels) in order to even better address the mentioned trade-off. This technical report is the first step needed to built identification experiment design for control in network context. 





\section*{Acknowledgment}

This work originally was supported by a grant from the R\'{e}gion Rh\^{o}ne-Alpes.

\bibliographystyle{IEEEbib}
\bibliography{IEEEabrv,gerard,anton_bib,DATABASE,identdir,biblio}

\end{document}

%% file: setup.tex
\renewcommand{\ni}{\noindent}

\newcommand{\bl}{\begin{list}{$\bullet$}{\topsep=1ex \parsep=.0ex
 \parskip=.0ex \itemsep=1ex}}
\newcommand{\el}{\end{list}}
\newcommand{\beqr}{\begin{eqnarray*}}
\newcommand{\eeqr}{\end{eqnarray*}}
\newcommand{\bs}{\begin{slide}}
\newcommand{\es}{\end{slide}}

\newcommand{\bdpm}{\begin{displaymath}}
\newcommand{\edpm}{\end{displaymath}}

\newcommand{\bi}{\begin{itemize}}
\newcommand{\ei}{\end{itemize}}

\newcommand{\ba}{\begin{array}} \newcommand{\ea}{\end{array}}

\newcommand{\be}{\begin{equation}}
\newcommand{\ee}{\end{equation}}

\newcommand{\beqna}{\begin{eqnarray}}
\newcommand{\eeqna}{\end{eqnarray}}

\newcommand{\bd}{\begin{displaymath}}
\newcommand{\ed}{\end{displaymath}}

\newcommand{\beqnd}{\begin{eqnarray*}}
\newcommand{\eeqnd}{\end{eqnarray*}}

\newcommand{\bvec}{\left ( \begin{array}}

\newcommand{\evec}{\end{array} \right )}
\newcommand{\bmtr}{\left[\begin{array}}
\newcommand{\emtr}{\end{array} \right] }

\renewcommand{\ni}{\noindent}

\newtheorem{theorem}{Theorem}[section]
\newtheorem{problem}{Problem}[section]

\newtheorem{definition}{Definition}[section]
\newtheorem{lemma}{Lemma}
\def\diss {$\left\lbrace x(\w),y(\w),z(\w)\right\rbrace$ - dissipative}
\def \sz  {\mathtt{s}}
\def \tom  {\varpi}
\def \dia {\textbf{diag}}

\def\BA  {\left[\begin{array}}

\def\EA{\end{array}\right]}
\def\BEQ {\begin{equation}}
\def\EEQ{\end{equation}}

\def \w     {\omega}

\def \jw    {j\omega}

\def \reals   {{\mbox{\bf R}}}
\def \eqbydef {\mathrel{\stackrel{\Delta}{=}}}

\def \cZ {{\mathcal Z}}

\def \cN {{\mathcal N}}
\def \cM {{\mathcal M}}

\def \cA {{\mathcal A}} 
\def \cB {{\mathcal B}}

\def \cY {{\mathcal Y}}

\def \cS {{\mathcal S}}

\def \cT {{\mathcal T}}

\def \cX {{\mathcal X}}

\def \pti {P_{\theta_i}}

\def \to {\theta_0}

\def \tho {\theta_0}

\def \thi {\hat{\theta}_i}
\def \tio {{\theta}_{i,0}}
\def \th {\hat{\theta}}
\def \to {{\theta}_{0}}

\def \Nm {N_{mod}}